\documentclass[fleqn,usenatbib]{mnras}

\usepackage{newtxtext,newtxmath}

\usepackage{ulem}

\usepackage[T1]{fontenc}

\DeclareRobustCommand{\VAN}[3]{#2}
\let\VANthebibliography\thebibliography
\def\thebibliography{\DeclareRobustCommand{\VAN}[3]{##3}\VANthebibliography}

\usepackage{graphicx}	

\makeatletter
\newcommand{\@todonotes@enable}{1}
\newcommand{\@todonotes@inline}{1}
\makeatother
\input{notes.tex}

\title[$\Lambda$CDM Substructure and Star Cluster Orbits]{The Effects of $\Lambda$CDM Dark Matter Substructure on the Orbital Evolution of Star Clusters}

\author[Pavanel and Webb]{
Nicholas Pavanel$^{1}$ and \thanks{E-mail: mn@ras.org.uk (KTS)}
Jeremy J. Webb$^{1}$
\\
$^{1}$Department of Astronomy and Astrophysics, University of Toronto, 50 St George St, Toronto, ON M5S 3H4, CA\\
}

\date{Accepted XXX. Received YYY; in original form ZZZ}

\pubyear{2021}

\begin{document}
\label{firstpage}
\pagerange{\pageref{firstpage}--\pageref{lastpage}}
\maketitle

\begin{abstract}

We present a comprehensive study on how perturbations due to a distribution of $\Lambda$CDM dark matter subhalos can lead to star clusters deviating from their orbits. Through a large suite of massless test particle simulations, we find that (1) subhalos with masses less than $10^8 M_{\odot}$ negligibly affect test particle orbits, (2) perturbations lead to orbital deviations only in environments with substructure fractions $f_{sub} \geq 1\%$, (3) perturbations from denser subhalos produce larger orbital deviations, and (4) subhalo perturbations that are strong relative to the background tidal field lead to larger orbital deviations. To predict how the variation in test particle orbital energy $\sigma_e(t)$ increases with time, we test the applicability of theory derived from single-mass subhalo populations to populations where subhalos have a mass spectrum. We find $\sigma_e(t)$ can be predicted for test particle evolution within a mass spectrum of subhalos by assuming subhalos all have masses equal to the mean subhalo mass and by using the local mean subhalo separation to estimate the change in test particle velocities due to subhalo interactions. Furthermore, the orbital distance variation at an orbital distance $r$ can be calculated via $\sigma_r=2.98 \times 10^{-5} \pm 8 \times 10^{-8} (\rm kpc^{-1} km^{-2} s^{2}) \times r \times \sigma_e$ with a dispersion about the line of best fit equalling 0.08 kpc. Finally, we conclude that clusters that orbit within 100 kpc of Milky Way-like galaxies experience a change no greater than $2\%$ in their dissolution times.  

\end{abstract}




\section{Introduction}

\label{sec:introduction}

The $\Lambda$ Cold Dark Matter ($\Lambda$CDM) framework explains the formation of structures like massive dark matter halos through the hierarchical merging of smaller subhalos \citep{white91,springel05}. Subhalos that reach dissolution due to external tides contribute to the smooth component of a galactic dark matter halo, while subhalos that have yet to disrupt exist as dark matter substructure. Galactic dark matter halos with a background 'smooth' potential component and a 'clumpy' substructure potential component are clearly found in $\Lambda$CDM simulations like The Aquarius Project (TAP) \citep{springel08} and Via Lactea II (VL-II) \citep{diemand2008}. TAP finds that Milky Way-like dark matter halos have radially dependant substructure mass fractions ($f_{sub}$), as they contain approximately 0.01\% of their total mass in substructure within the solar circle, an upper limit of 3\% within 100kpc, and an upper limit of 20\% within 400kpc. VL-II finds subhalos to have a mass-size relationship given by $1.05kpc\left(\frac{M_{Sub}}{10^{8}M_{\odot}}\right)^{\frac{1}{2}}$, assuming that subhalos are well represented as Hernquist spheres.

State-of-the-art cosmological simulations, which also work to include the effects of baryonic matter on dark matter subhalo evolution, have attempted to refine these estimates. However there is scatter in predicted values of $f_{sub}$ due to resolution effects and the details in which the baryonic component is modelled \citep{DOnghia10, Zolotov12, Brooks13, Brooks14,wetzel16, Sawala17, garrisonkimmel17,kelley19, Richings20, Webb2020, Grand2021}. Most recent theoretical studies agree, however, that dark matter-only simulations overestimate the number of subhalos in Milky Way-like galaxies.

To validate the $\Lambda$CDM cosmological model, a significant amount of work has been devoted to observationally verifying the existence of dark matter subhalos. The majority of studies have focused on the effects of subhalos on gravitational lensing and stellar streams. Dark matter substructure has been shown to produce distortions in gravitationally lensed objects \citep{mao98}, allowing for Galactic subhalos with masses ranging between $10^6M_{\odot}$ and $10^8M_{\odot}$ to be indirectly observed \citep{gilman2020}. These measurements of dark matter subhalo masses place large dark matter subhalos into the mass range of sub-dwarf galaxies.

With respect to stellar streams, \citet{yoon2011} used numerical techniques to show that a population of dark matter subhalos can leave observable surface density fluctuations in cold stellar streams such as GD1 and Pal 5. Underdensities \citep{bovy17} and spurs \citep{bonaca19} have become telltale features that observers look for in stellar streams to infer a dark-matter subhalo interaction has taken place (see also \citet{carlberg2012, carlberg2013, erkal2015_1, erkal2015_2, sanders2016}). A close analysis of these features further allows for the properties of subhalos to be constrained, with \citet{bonaca19} suggesting a subhalo could only produce the GD1 spur if it was 10 times denser than a standard $\Lambda$CDM subhalo.

It has not been until recently that applying the theoretical works above to observational data has led to observational constraints being placed on $f_{sub}$. \citet{banik2019} determined that gaps in stellar streams like GD1 and PAL5 cannot be explained by baryonic matter alone, but can be explained when subhalo effects are included. The study also placed a constraint on the Milky Way's substructure mass fraction at $0.14^{+0.11}_{-0.07}\%$ within 20kpc. The work done by \citet{banik2019} is, to date, the only observational measurement that constrains the Milky Way's substructure mass fraction. Hence further work is needed to confirm the existence of substructure and determine its properties.

Additional methods are required to constrain the properties of dark matter substructure in the Milky Way in order to compliment studies of density perturbations along stellar streams. For example, there is evidence to suggest that massive subhalos are capable of peturbing the orbits of not only stellar streams, but globular clusters and satellite galaxies as well. The orbits of satellite galaxies have been shown to be sensitive to interactions with the Magellanic Clouds \citep{Patel20} and to how the Milky Way's halo responds to interactions with other satellites \citep{GaravitoCamargo19}. Furthermore, studies such as \citet{gomez2015}, \citet{erkal2018}, and \citet{erkal2019} have shown that the Large Magellanic Cloud (LMC) is able to perturb stars in the tidal tails of the Sagittarius dwarf galaxy, the Tuscan stellar stream, and the Orphan stellar stream such that their velocities are misaligned with respect to their stream paths. With respect to globular clusters, \citet{Garrow} finds that interactions with dwarf galaxies between the masses of $10^9M_{\odot}$ and $10^{11}M_{\odot}$ can cause the orbital properties of Galactic globular clusters to evolve significantly over time. 

While encounters between stellar systems and dark matter subhalos will be more numerous than encounters with satellite galaxies, most subhalos will have masses less than $10^9M_{\odot}$. Thus their gravitational influence on a stream or globular cluster will be weaker. Hence whether or not dark matter subhalos are capable of perturbing the orbits of stellar systems as much as satellite galaxies remains to be determined.

\citet{Penarrubia2019} studied the perturbative effects that extended substructure has on massless test particles and found that random interactions with extended substructure are capable of scattering test particle orbits. The author found that the variation in a particle's orbital energy $\sigma_E$ grows with time (t) as a function of the ratio of the mean separation (D) and size (c) of the substructure as well as the mean lifetime of perturbative forces ($T_{ch}$). More specifically, \citet{Penarrubia2019} derives that $\sigma_E^2 \propto <v^{-2}> {t/T_{ch}}^2$ for $t<<T_{ch}$ and $\sigma_E^2 \propto <v^{-2}> {t/T_{ch}} \ln{C/d}$ for $t>>T_{ch}$, where $<v^{-2}>$ is the mean square speed of the subhalos. Since star clusters are some of the oldest structures in the Milky Way \citep{krauss2003, MarinFranch09, Forbes10}, they will have tidal histories that are comparable to the test particles in \citet{Penarrubia2019} in that they will have experienced a large number of subhalo interactions over the course of their lifetimes. Is is therefore possible that interactions with subhalos have affected the orbital evolution of individual star clusters. 

\cite{Webb2019} determined that tidal interactions with CDM subhalos with masses between $10^5M_{\odot}$ and $10^{11}M_{\odot}$ approximated as Hernquist spheres are too weak to dissolve star clusters via tidal heating directly. However, this study did not account for the ability of subhalos to altar a cluster's orbital path. Given how strongly globular cluster evolution is linked to its orbit in an external tidal field \citep{baumgardt2003, Webb2015}, it is possible that deviations in a cluster's orbit will produce unique effects in the perturbed cluster's mass, structure, and stellar mass function.

Motivated by \citet{Penarrubia2019}, this study focuses on how $\Lambda$CDM substructure affects the orbital evolution of test particles on star cluster-like orbits. We specifically focus on how these effects depend on subhalo mass, subhalo size, and the overall substructure mass fraction of a galaxy. This study will help determine the range of subhalo masses and sizes that need to be included in cosmological simulations that wish to accurately model the orbital evolution of star cluster particles. It will also extend the work of \citet{Penarrubia2019} to include substructure populations with a mass spectrum. In Section \ref{sec:methods} we outline how we initialize test particle orbits in analytic potentials containing substructure. Section \ref{sec:results} will demonstrate how both the orbital energy and radius of the test particles vary as a function of time due substructure interactions, and how this variance depends on subhalo mass, size, and mass fraction. In Section \ref{sec:discussion} we apply the work of \citet{Penarrubia2019} to the orbital energy variance evolution of the test particles and introduce an analytic fit for how orbital distance varies with time. We also consider the implications these variations have on star cluster evolution and then summarize our findings in Section \ref{sec:conclusion}.

\section{Methods}

\label{sec:methods}

In order to study how $\Lambda$CDM substructure can affect the orbital evolution of star clusters, it is necessary to integrate the orbits of massless test particles orbiting in dark matter halos with varying amounts of substructure and substructure properties. Differences between how test particles orbit in potentials with and without substructure will demonstrate how the orbits of star clusters can change over a lifetime of substructure interactions. The properties of a galaxy's substructure population, including the substructure fraction $f_{sub}$, the subhalo mass function, and individual subhalo densities will play a role in how strongly cluster orbits are affected.  

We start by assuming a simple base galaxy model with no substructure that takes the form of a logarithmic potential with a circular velocity of $\rm 220 \ {km / s^{-1}}$ at $\rm 8 \ kpc$. This potential is referred to as the total potential, and is analogous to the Galactic dark matter halo \citep{xue2008}. Galaxy models with smooth and substructure components are also taken to be logarithmic potentials, where the amplitude of total potential is scaled by (1-$f_{sub}$) to represent the smooth component and by $f_{sub}$ to represent the substructure component. In this study, we consider galaxy models with $f_{sub}$ values of $0.1\%$, $1\%$, $3\%$, and $10\%$.

In order to generate subhalo populations based on the substructure component of our potential, we first generate subhalo masses. To explore how subhalos of different masses can perturb test particle orbits, we first consider single mass subhalo populations with masses of $10^6M_{\odot}$, $10^7M_{\odot}$, $10^8M_{\odot}$ and $10^9M_{\odot}$ in galaxy models with substructure fractions of $3\%$ (Models M6f3, M7f3, M8f3, M9f3). Subhalos with masses below $10^5M_{\odot}$ have been shown to have no observable effects on stellar stream density \citet{banik2019}, and we assume that their perturbative effects on circular test particle orbits will also be negligible. We also consider the more realistic case of subhalo populations with a spectrum of masses, where we choose subhalo masses between $10^6M_{\odot}$ and $10^9M_{\odot}$ based on a mass function of the form $\frac{dN}{dM} \propto M^{-2}$ \citet{moore1999,klypin1999,diemand2008,springel08}. We primarily consider subhalos as Hernquist spheres with the mass-radius relationship consistent with \citet{diemand2008, springel08, erkal2016} and equal to: 

\begin{equation}
\centering
r_{s} = 1.05kpc\left(\frac{M_{Sub}}{10^{8}M_{\odot}}\right)^{\frac{1}{2}}.
\label{eqn:subhalomass_radius}
\end{equation}

where $r_s$ is the Hernquist sphere scale radius. Subhalo populations with a full range of masses are generated for $f_{sub}$ values of $0.1\%$, $1\%$, $3\%$, and $10\%$ (Models MSf01,MSf1,MSf3,MSf10). Finally, we also consider one case where subhalo scale radii are $10\times$ smaller than Equation \ref{eqn:subhalomass_radius} would suggest, a possibility suggested by \citet{bonaca19} to explain the the GD1 spur (Model MSf3r). Given the resolution limits of cosmological simulations like TAP and VL-II, it is possible that dark matter substructure has a different mass-size relationship. 

The properties of each potential model are summarized in Table \ref{table:sim}. Model names are given such that the character after the initial M marks either the logarithm of the subhalo mass (in single mass models) or an S if a mass spectrum is used. The number after the f marks the substructure fraction. Hence a galaxy model where subhalos have a mass spectrum and a substructure fraction of $3\%$ has a name of MSf3. In one case, a trailing r in the model name marks that subhalos are $10 \times$ smaller than \ref{eqn:subhalomass_radius}.

For a given potential model, subhalo positions and velocities are initialized so the population reflects the substructure logarithmic potential out to 100 kpc. Positions are randomly sampled from the logarithmic potential's corresponding density profile and velocities are sampled from a Gaussian distribution with a dispersion equal to $\frac{v_{circ}}{\sqrt{3}}$, where $v_{circ}$ is the circular velocity at the subhalo's galactocentric distance. For each model we calculate the mean $C/D$ and $T_{ch}$ and list the values in Table \ref{table:sim} in order to easily compare with \citet{Penarrubia2019}. Using the galactic dynamics package  \texttt{galpy} \footnote{http://github.com/jobovy/galpy} \citet{galpy2015} the distribution of subhalos then have their orbits integrated in the total potential. More specifically, each subhalo is individually integrated in the total potential; orbiting subhalos do not influence each other. This process is repeated $100 \times$ per galaxy model, such that we have 100 realizations of a galaxy with a given substructure mass function, mass fraction, and mass-size relationship. For each subhalo distribution we integrate five test particles at initial galactocentric radii of  5 kpc, 10 kpc, 20 kpc, 40 kpc, and 60 kpc for 12 Gyr (the mean age of Galactic globular clusters \citep{forbes2010}).

\begin{table*}
\centering
\begin{tabular}{c|c|c|c|c|c|c|c|}
\hline
Name & Subhalo Mass Range & Substructure Mass Fraction & Mass-Size relation & $C/D$ & $T_{ch}$ \\ [0.5ex] 
\hline\hline
{M6f3} & $10^6M_{\odot}$ & 3\% & Eqn. \ref{eqn:subhalomass_radius} & 0.04 & 0.02 Gyr \\
\hline
{M7f3} & $10^7M_{\odot}$ & 3\% & Eqn. \ref{eqn:subhalomass_radius} & 0.06 & 0.04 Gyr \\
\hline
{M8f3} & $10^8M_{\odot}$ & 3\% & Eqn. \ref{eqn:subhalomass_radius} & 0.09 & 0.08 Gyr \\
\hline
{M9f3} & $10^9M_{\odot}$ & 3\% & Eqn. \ref{eqn:subhalomass_radius} & 0.14 & 0.16 Gyr \\
\hline
{MSf01} & $10^6M_{\odot} - 10^9M_{\odot}$ & 0.1\% & Eqn. \ref{eqn:subhalomass_radius} & 0.01 & 0.1 Gyr \\
\hline
{MSf1} & $10^6M_{\odot} - 10^9M_{\odot}$ & 1\% & Eqn. \ref{eqn:subhalomass_radius} & 0.03 & 0.05 Gyr \\
\hline
{MSf3} & $10^6M_{\odot} - 10^9M_{\odot}$ & 3\% & Eqn. \ref{eqn:subhalomass_radius} & 0.05 & 0.03 Gyr \\
\hline
{MSf10} & $10^6M_{\odot} - 10^9M_{\odot}$ & 3\% & Eqn. \ref{eqn:subhalomass_radius} & 0.07 & 0.02 Gyr \\
\hline
{MSf3r} & $10^6M_{\odot} - 10^9M_{\odot}$ & 3\% & Eqn. \ref{eqn:subhalomass_radius}/10. & 0.01 & 0.02 Gyr \\
\hline
\end{tabular}
\caption{Name and properties of each galaxy model. Model names reflect whether or not a single subhalo mass or mass spectrum is used, the substructure mass fraction, and the subhalo mass-size relationship.}
\label{table:sim}
\end{table*}

\begin{figure}
\centering
\includegraphics[width=1.0\linewidth]{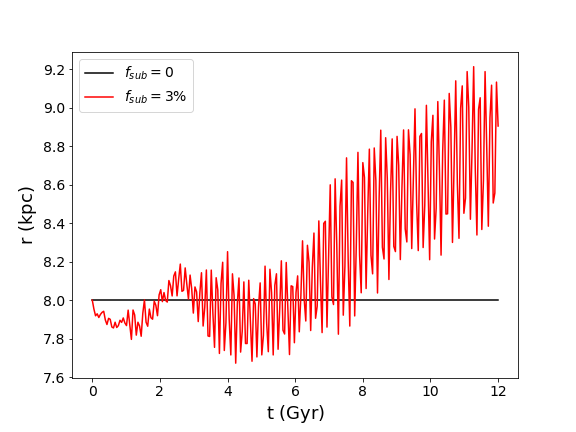}
\caption{Galactocentric distance as a function of time for a single test particle initially with a circular orbit at 4kpc in the smooth galaxy model (black) and the M8f3 galaxy model (red). Subhalo interactions result in significant fluctuations in the test particle's orbit.}
\label{fig:one_test particle_evolution}
\end{figure}


\section{Results}

\begin{figure}
\centering
\includegraphics[width=1.0\linewidth]{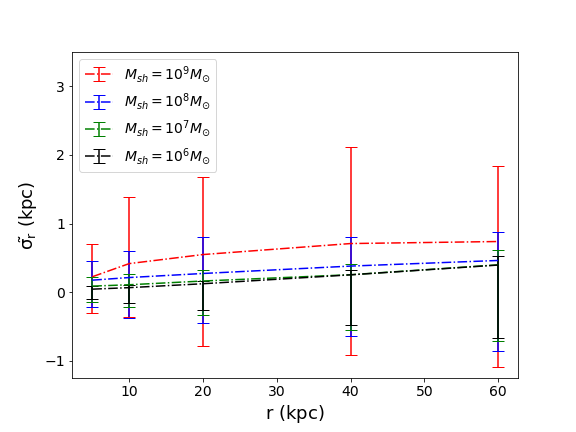}
\caption{Median dispersion $\sigma_r$ in galactocentric distance for test particles with initially circular orbits at 5, 10, 20, 40, and 60 kpc orbiting in galaxy models M6f3, M7f3, M8f3, and M9f3. Each dispersion is calculated from 100 realizations of the galaxy model. Error bars represent represent the median maximal distance away from initialization that each test particle reached. $\sigma_r$ increases as a function of both orbital distance and subhalo mass, implying that fewer encounters with higher-mass subhalos result in larger perturbations and orbital deviations than many encounters with lower-mass subhalos.}
\label{fig:mass_study}
\end{figure}

\begin{figure}
\centering
\includegraphics[width=1.0\linewidth]{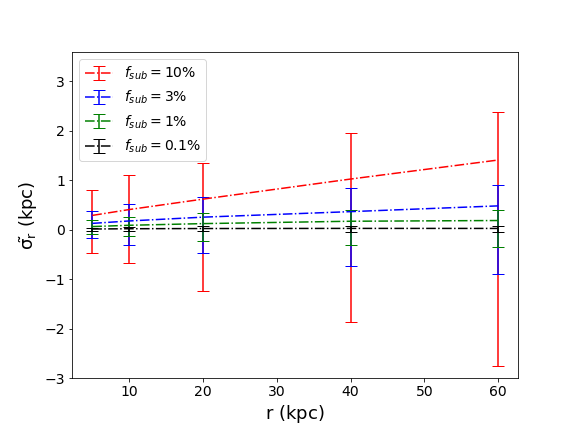}
\caption{Same as Figure \ref{fig:mass_study}, but for test particles orbiting in galaxy models MSf01, MSf1, and MSf3.}
\label{fig:smf_study}
\end{figure}

\label{sec:results}

To illustrate how our analysis works, we first consider in Figure \ref{fig:one_test particle_evolution} the evolution of a single test particle in a potential without substructure and in a potential with $10^{8}M_{\odot}$ subhalos and $f_{sub} = 3\%$. The initial orbit is circular in the smooth potential, which is why the radial distance of the unperturbed test particle stays constant. Including substructure leads to perturbations that cause the test particle to deviate from its orbit. To quantify the deviation for this individual case we calculate the standard deviation in the test particle's galactocentric radius, as well as the minimum and maximum galactocentric radii. For the particle in Figure \ref{fig:one_test particle_evolution}, these values are 0.12 kpc, 7.78 kpc, and 9.27 kpc. We then repeat the orbital integration in one hundred different realizations of the potential with substructure and calculate the median standard deviation $\sigma_r$, and median minimum $r_{min}$ and maximum $r_{max}$ galactocentric distance values. Exploring how these values depend on galactocentric distance, $f_{sub}$ the subhalo mass function, and the subhalo mass-radius relation allow for an in-depth analysis on how substructure interactions affect cluster orbits. 

\subsection{Dependence on Subhalo Mass}

\label{subsec:mass_study}

We begin our analysis by exploring how strongly subhalos with different masses perturb test particle orbits. Figure \ref{fig:mass_study} illustrates $\sigma_r$ for test particles initially with circular orbits at 5, 10, 20, 40, and 60 kpc in galaxy models M6f3, M7f3, M8f3, and M9f3. The upper and lower error bars represent $r_{max}$ and $r_{min}$ respectively. In general $\sigma_r$ increases from 0.04 kpc $\leq \sigma_{r} \leq 0.22$ kpc in the inner regions to 0.39 kpc $\leq \sigma_{r} \leq 0.74$kpc in the outer regions, with galaxy model M9f3 yielding the largest $\sigma_r$ at 60 kpc of 0.74 kpc. In the most extreme cases, test particle orbits can reach distances up to 2 kpc away from their original orbit.

Figure \ref{fig:mass_study} displays two important trends. Firstly, $\sigma_r$, $r_{max}$ and $r_{min}$ all increase with subhalo mass for a given orbital distance. Hence fewer perturbations from high-mass subhalos results in larger orbital deviations than many perturbations from low-mass subhalos. Secondly, $\sigma_r$, $r_{max}$ and $r_{min}$ all increase with orbital distance for a given subhalo mass. Therefore the strength of a subhalo perturbation relative to the smooth tidal field is also of importance. Test particles orbiting at large galactocentric distances, where the smooth tidal field is weak, can more easily deviate from their orbital path when perturbed by a subhalo.

\subsection{Dependence on Subhalo Mass Fraction}

\label{subsec:smf_study}

We next consider how strongly test particle orbits can be perturbed in galaxy models with different substructure mass fractions $f_{sub}$. Figure \ref{fig:smf_study} illustrates $\sigma_r$ for test particles initially with circular orbits at 5, 10, 20, 40, and 60 kpc in galaxy models MSf01, MSf1, MSf3, and MSf10. It should be noted that the subhalos in these galaxy models have a range of masses that reflect a power law of slope -2 between $10^6$ and $10^9 M_{\odot}$. Hence these simulations consist of subhalos that are similar to $\Lambda$CDM simulations like TAP and VL-II. Furthermore, our 0.1\% substructure mass fraction simulation produces results that can be applied to the Milky Way as \citet{banik2019} measured the substructure mass fraction to be $0.14^{+0.11}_{-0.07}\%$ within 20kpc.

Several important trends are revealed when analyzing Figure \ref{fig:smf_study}. Firstly, for $f_{sub}=0.1\%$, the orbits of tracer particles are minimally affected by the presence of substructure. In the MSf01 model, tracer particles have a mean $\sigma_r$ of $0.023$ kpc $\pm 0.002$, with $r_{min}$ and $r_{max}$ reaching at most 0.05 kpc. Secondly, taking into consideration larger values of $f_{sub}$, $\sigma_r$, $r_{max}$ and $r_{min}$ generally increase with substructure mass fraction. More specifically, $\sigma_r$ is 5.6, 12.4, and 32.9 times larger when $f_{sub} = 1\%$, $3\%$, and $10\%$ compared to the $0.1 \%$ case. This result is not surprising, as the subhalo encounter rate increases with $f_{sub}$. Hence for galaxy models where $f_{sub}$ increases as a function of galactocentric distance, as seen in TAP and VL-II, outer globular clusters will experience stronger perturbations and larger orbital deviations than inner clusters. Finally, similar to Figure \ref{fig:mass_study}, $\sigma_r$, $r_{max}$ and $r_{min}$ all increase with orbital distance. This behaviour supports our previous claim that the relative strength of subhalo perturbations is stronger at larger radii, resulting in larger orbital deviations.

The MSf3 models allow for an exploration of how differently orbits are affected by a single-mass population of subhalos compared to a subhalo population with a mass spectrum. More specifically, $\sigma_r$ is 0.5 times smaller in the MSF3 model than in M9f3, 0.9 times smaller than in M8f4, 1.4 times larger than in M7f3, and 1.60 times larger than in the M6f3. These ratios suggest that a subhalo population with masses between $10^6$ and $10^9 M_{\odot}$ that follow a power law distribution function with a slope of -2 behave like a single-mass subhalo population with masses between $10^7$ and $10^8 M_{\odot}$ despite the mean subhalo mass in the MSf3 model being $\sim 7 \times 10^6 M_{\odot}$. Hence for a given $f_{sub}$, the increased number of subhalos in a population that follows a mass spectrum partially compensates for the decreased number of high-mass subhalo interactions by introducing a larger number of low-mass interactions. Comparing this result to Section \ref{fig:mass_study}, where we find fewer high mass subhalo interactions lead to larger orbital perturbations compared to more low mass subhalo interactions, it appears that the low mass subhalo interactions still play an important role in orbital evolution when combined with high mass subhalo interactions.

\subsection{Dependence on Subhalo Density} \label{subsec:density_study}

\begin{figure}
\centering
\includegraphics[width=1.0\linewidth]{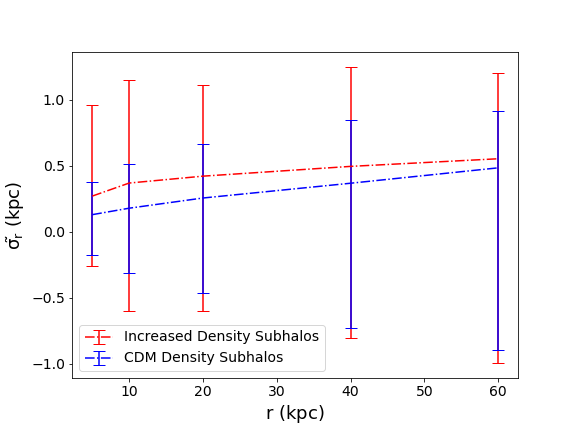}
\caption{Same as Figure \ref{fig:mass_study}, but for test particles orbiting in galaxy models MSf3 and MSf3r.}
\label{fig:density_study}
\end{figure}

Finally, we test how strongly particle orbits can be perturbed in galaxy models with subhalos that are denser than the standard $\Lambda$CDM model. Figure \ref{fig:density_study} illustrates $\sigma_r$ for test particles initially with circular orbits at 5, 10, 20, 40, and 60 kpc in galaxy models MSf3 and MSf3r, where subhalos are $10\times$ denser in MSf3r. 

We expectedly find that $\sigma_r$, $r_{max}$ and $r_{min}$ are higher in the galaxy model with denser subhalos, as interactions with compact subhalos result in stronger perturbations than interactions with extended subahlos. The difference appears to be more significant in the inner regions where the encounter rate is high, with $\sigma_r$ being approximately a factor of 2 larger in MSf3r for orbital distances less than 20 kpc. At larger distances the factor decreases to 1.2.

\section{Discussion} \label{sec:discussion}

\begin{figure*}
\centering
\includegraphics[width=\textwidth]{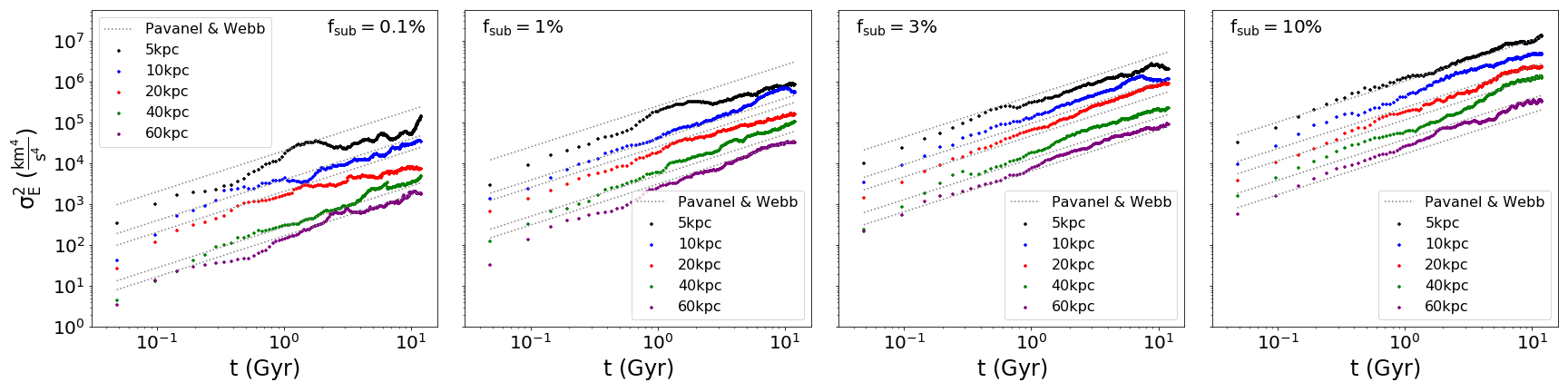}
\caption{Standard deviation in orbital energy $\sigma_E$ of all test particles as a function of time for all particles orbiting at 5, 10, 20, 40 and 60 kpc in galaxy models MSf01 (left panel), MSf1 (centre-left panel), MSf3 (centre-right panel) and MSf10 (right panel). $\sigma_E$ increase as a function of time as particles undergo repeated interactions with subhalos, with $\sigma_E$ also increasing with $f_{sub}$ and decreasing with orbital distance. Theoretical predictions from \citep{Penarrubia08} are illustrated for comparison purposes}
\label{fig:e_vari}
\end{figure*}

We have explored how perturbations by dark matter subhalos can alter the orbits of massless test particles, with a particular focus on how orbital deviations depend on subhalo mass, a galaxy's substructure mass fraction, and the subhalo mass-radius relationship. In general, we find that larger perturbations and orbital deviations are produced from subhalos distributions with a small number of high mass subhalos than from distributions with a large number low mass subhalos. Hence the strength of encounters, not the frequency of encounters, is the dominant subhalo property. Also playing a role here is that, since weak encounters are more frequent, an isotropic distribution of encounters will lead to the particles orbit remaining approximately constant. Since stronger encounters are rarer, its more likely that the distribution of strong encounters is anisotropic over the timescales considered here. However when strong and weak encounters are occurring due to subhalos having a mass spectrum, low mass subhalo interactions can lead to larger orbital perturbations than the mean mass of the subhalo population would suggest.

Secondly, perturbations lead to deviations only when the substructure mass fraction is larger than or equal to 1\%. While a trend with $f_{sub}$ is not surprising, since a larger $f_{sub}$ leads to more subhalo interactions, the fact that particle orbits are minimally affected for $f_{sub} \le 0.1 \%$ means that most inner region Galactic globular clusters will remain unaffected by subhalo interactions given the \citet{banik19} estimate of $f_{sub} = 0.14^{+0.11}_{-0.07}\%$ within 20kpc. Thirdly, denser subhalos expectedly lead to stronger orbital perturbations and larger orbital deviations since the gravitational force of more compact subhalos is stronger than extended subhalos. Finally, as shown throughout all simulations, the strength of a given perturbation relative to the background smooth tidal field determines how strongly a test particle's orbit can deviate from its initial path. More specifically, for a given perturbation strength, if the test particle is located in the inner regions of the galaxy then the relative perturbation strength is low and it will be difficult to alter the particle's orbit. Conversely, if the particle is located in the outskirts of the galaxy where the smooth tidal field is weak, its much easier for the perturbation to lead to the particles orbital path being changed.

In the following subsections, we first work to understand these conclusions by exploring how the change in orbital radius is related to changes in orbital energy with time. By relating changes in orbital radius with time to orbital energy we are able to directly compare our findings to \citet{Penarrubia2019}, who provides a theoretical estimate for how the distribution of orbital energies should grow with time for a given population of subhalos. We then extrapolate these results to provide a theoretical estimate for how the distribution of orbital radii increases with time and relate the results to how globular cluster dissolution times are affected by the change in orbital distance.

\subsection{Evolution of dispersion in orbital energies}

\begin{figure*}
\centering
\includegraphics[width=\linewidth]{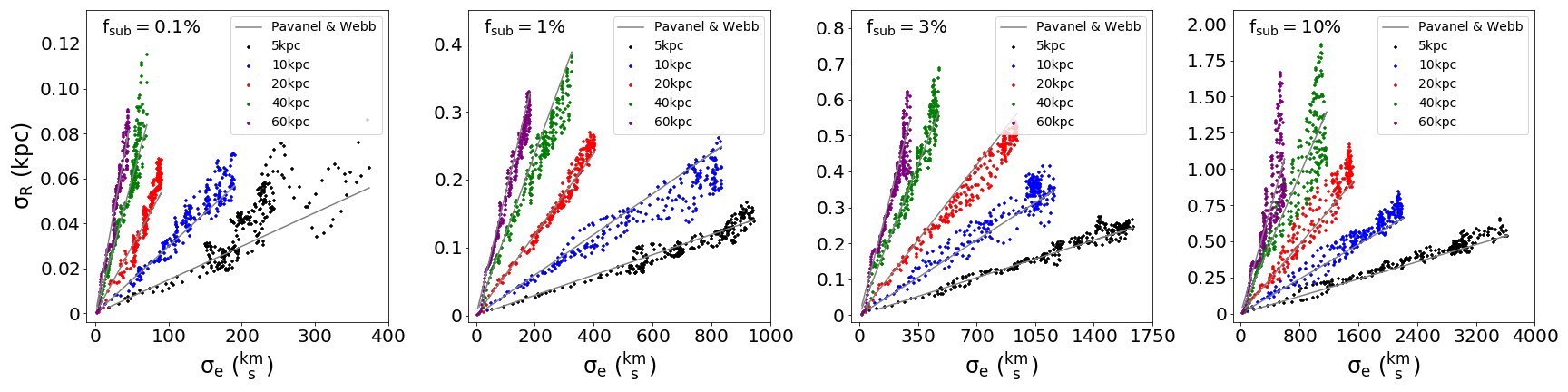}
\caption{Standard deviation in orbital distance $\sigma_r$ of all test particles as a function of time for all particles orbiting at 5, 10, 20, 40 and 60 kpc in galaxy models MSf01 (left panel), MSf1 (centre-left panel), MSf3 (centre-right panel) and MSf10 (right panel). $\sigma_r$ increase as a function of time as particles undergo repeated interactions with subhalos, with $\sigma_r$ also increasing with $f_{sub}$ and orbital distance $r$. The fit to each model, using Equation \ref{eqn:sigr_fit}, is also illustrated.}
\label{fig:sige_sigr}
\end{figure*}

The underlying assumption in our analysis is that test particles orbit within their respective model galaxy for 12 Gyr. For each simulation, test particle-subhalo interactions happen at different points of the test particle's evolution and are therefore a time-dependant phenomenon. Hence one can imagine that for test particles orbiting for less than 12 Gyr, all reported values of $\sigma_r$ will decrease and the difference between each particle's initial radius and both $r_{max}$ and $r_{min}$ will decrease as well. It is therefore worthwhile to consider the time dependence of $\sigma_r$ in order to understand the timescale over which subhalo interactions are important.

\citet{Penarrubia2019} explored the variation in orbital energy that an individual test particle has as a function of time in simulations with different substructure properties. \citet{Penarrubia2019} found that the orbital energy of test particles that evolve in an environment with an abundance of extended substructures increases quadraticly with time t when $t<<T_{ch}$ and linearly with time when $t>>T_{ch}$. We perform a similar analysis in Figure \ref{fig:e_vari}, where each data point corresponds to the standard deviation in the orbital energy values $\sigma_E$ that test particles with the same initial orbital distance have at each timestep. In all cases, $\sigma_E$ increases at a near linear rate as a function of time as test particles undergo repeated interactions with subhalos. Therefore, the longer a particle spends in a potential with extended substructures, the more it will interact with substructure and experience changes in its orbital energy.

To compare directly with \citet{Penarrubia2019}, we make use of Equations 18 and 29 in \citet{Penarrubia2019} to calculate the velocity increment $\Delta{\mathbfit v}$ experienced by a particle orbiting in a population of single-mass subhalos of mass $M$, size $c$, and mean separation $D$ via

\begin{equation}\label{eq:delv2}
  \langle |\Delta{\mathbfit v}|^2\rangle \approx t\,\sqrt{\frac{8\pi}{3\langle v^2\rangle}}\frac{(GM)^2}{D^3}\big[\ln(D/c)-1.9\big]
\end{equation}

and the corresponding increase in the orbital energy dispersion $\sigma_e$ with time via

\begin{equation}\label{eq:delE2}
  \sigma_E^2(t)=\frac{1}{3}\overline{v_\star^2}\langle |\Delta{\mathbfit v}|^2\rangle 
\end{equation}

\noindent where $v_\star^2$ is the circular orbit velocity at the particle's orbital distance. While Equations \ref{eq:delv2} and \ref{eq:delE2} have been tested against simulations by \citet{Penarrubia2019}, the simulations consisted of single-mass subhalo populations and particles orbiting at distances comparable to D in a \citet{Dehnen93} potential. To compare the theoretical predictions to subhalos with a mass spectrum orbiting over a range of distance in a logarithmic potential, we take the mean subhalo mass for $M$, the predicted size of a subhalo of mass M given Equation \ref{eqn:subhalomass_radius} for $c$, and the local mean separation between subhalos based on a particle's orbital distance to be D. 

As illustrated in Figure \ref{fig:e_vari}, to first order Equations \ref{eq:delv2} and \ref{eq:delE2} are able to accurately predict $\sigma_{e}(t)$. Hence the derivations made by \citet{Penarrubia2019} also apply to subhalo populations with a mass spectrum and particles orbiting at different galactocentric distances as long as the mean subhalo mass M and local D are used. The dispersion of the residual between the true values and the fit values are 57 $\frac{km^2}{s^{2}}$ for MSf01, 225 $\frac{km^2}{s^{2}}$ for MSf1, 217 $\frac{km^2}{s^{2}}$ for MSf3, and 202 $\frac{km^2}{s^{2}}$ for MSf10. The minor offsets between \citet{Penarrubia2019} and our simulations are the result of Equations \ref{eq:delv2} and \ref{eq:delE2} overestimating $\sigma_{e}$. This result is surprising as we naively expected using the mean subhalo mass $<M>$ of the population for M would underestimate $\sigma_{e}$ based on our finding that subhalo populations with a mass spectrum yield $\sigma_{r}$ values that are larger then single mass subhalo populations with masses equal to $<M>$. The offsets must therefore stem from the fact that the \citet{Penarrubia2019} derivation assumes orbital distances comparable to D, while we explore a range of orbital distances and are forced use the mean local D which can vary over time.

\subsection{Evolution of dispersion in orbital radii}

\begin{figure*}
\centering
\includegraphics[width=\linewidth]{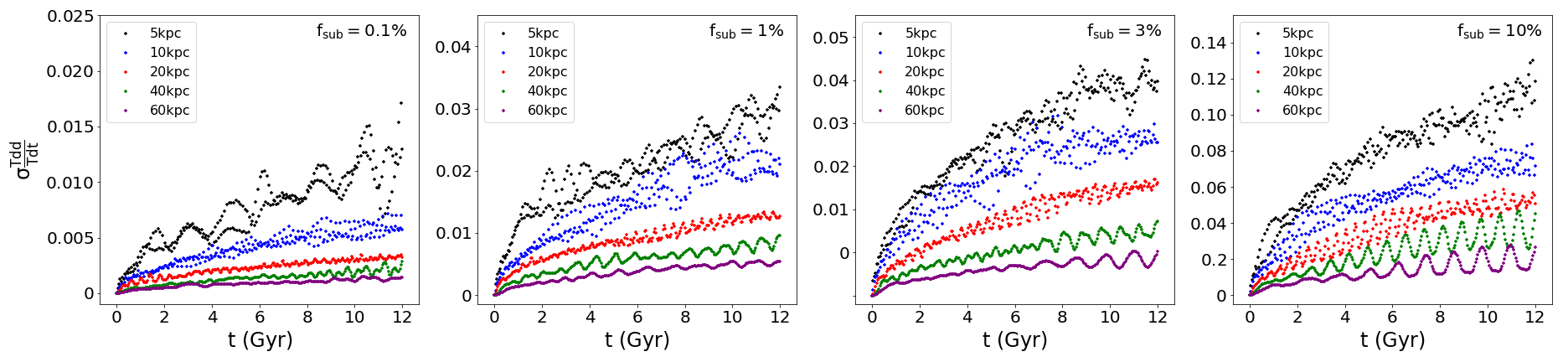}
\caption{Standard deviation in the ratio of theoretical cluster dissolution times in galaxy models with substructure to dissolution times in galaxy models with no substructure $\sigma_{T_{diss}}$ for all test particles as a function of time for all particles orbiting at 5, 10, 20, 40 and 60 kpc in galaxy models MSf01 (left panel), MSf1 (centre-left panel), MSf3 (centre-right panel) and MSf10 (right panel). $\sigma_{T_{diss}}$ increases as a function of time as repeated interactions with subhalos cause cluster orbital distances to vary, with $\sigma_{T_{diss}}$ reaching as high as $14\%$ for large values of $f_{sub}$ and small orbital distances.}
\label{fig:sigtdiss}
\end{figure*}

A similar analysis to the one above can also be performed based on how the standard deviation in each test particles galactocentric radius $r$ varies with time. Figure \ref{fig:sige_sigr} illustrates the relationship between $\sigma_r$ and $\sigma_e$, which appears to be nearly linear with the slope of the relationship depending on the test particle's orbital distance $r$. To quantify this relationship, we fit all the data in Figure \ref{fig:sige_sigr} with a relation of the form:

\begin{equation}\label{eqn:sigr_fit}
\sigma_r=A \times r \times \sigma_e
\end{equation}

\noindent Hence we assume $\sigma_r$ increases linearly with $\sigma_e$, with the rate of increase linearly depending on the particles orbital distance. We find the best fit value of A to be $2.98 \times 10^{-5} \pm 8 \times 10^{-8} (\rm kpc^{-1} km {-2} s^{2})$. The dispersion of the residual between the true values and the fit values is 0.08 kpc. The low dispersion indicates that the relationship between $\sigma_r$ and $\sigma_e$ is well represented by a linear function. The best fit line for each model is also illustrated in Figure \ref{fig:sige_sigr}.

The linear relationship between $\sigma_r$ and $\sigma_e$ suggests that the dispersion in orbital energies is primarily driven by subhalo encounters perturbing test particles to new orbital distances. Scatter about the relationship is likely due to test particles being pushed to orbits with a range of eccentricities as opposed to new circular orbit distances. The fact that the scaling parameter A is independent of orbital distance and $f_{sub}$ supports our finding that it is the relative strength of subhalo perturbations, compared to the background smooth tidal field, that determines how strongly a test particle's orbit will deviate from its initial trajectory. At small orbital distances, energetic perturbations are less capable of altering a test particles orbital distance relative to the same perturbation occurring at a large orbital distance. 

\subsection{Globular cluster dissolution times}

Motivated by findings that globular cluster evolution within a tidal field is governed strongly by orbital distance \citep{baumgardt2003, Webb2015}, we explore how a cluster's dissolution time is affected by subhalo induced perturbations and orbital deviations. \citet{baumgardt2003} estimate the dissolution times of a cluster moving on an orbit with eccentricity $\epsilon$ through a logarithmic potential via:

\begin{equation}\label{eq:tdiss1}
    T_{Diss} = \beta \bigg(\frac{N}{ln(0.02 N)}\bigg)^{x} \frac{R}{V} (1 - \epsilon) ,
\end{equation}

where $\beta$ and $x$ depend on the cluster's structural properties. For two identical clusters with different orbital distances, to quantify how strongly subhalo interactions can affect a cluster's dissolution time, we consider the dispersion of dissolution times that could result from subhalo interactions for each test particle. More specifically, we make use of how $\sigma_{R}$ evolves with time for each test particle and sample 1000 orbital distance values at each timestep from a Gaussian distribution centered at each test particle's unperturbed radius. Given Equation \ref{eq:tdiss1}'s linear dependence on R and the constant circular orbit velocity of a logarithmic potential, if we naively assume the tracer particles are perturbed to circular orbits then the ratio of each randomly generated radius to the test particle's unperturbed radius represents the ratio of dispersion times $\sigma_{T_{diss}}$ that two identical clusters orbiting at these radii would have. 

In Figure \ref{fig:sigtdiss} we plot the standard deviation of these dissolution time ratios at each timestep for each test particle. Thus, Figure \ref{fig:sigtdiss} represents the dispersion in the fractional change of the dissolution time for a cluster at each particle's respective radius due to subhalo induced perturbations. The estimates for $\sigma_{T_{diss}}$ are effectively lower limits, as we have not considered the effect of subhalo perturbations increasing the orbital eccentricity of the test particles.

The leftmost panel of Figure \ref{fig:sigtdiss} shows that clusters that evolve in potentials with $f_{sub} = 0.1\%$ have dissolution times that vary by at most $1.5\%$ from the dissolution times of clusters in potentials without substructure. This result is expected because potentials with $f_{sub} = 0.1\%$ host a small population of subhalos, and because we find negligible orbital perturbations and deviations regardless of the test particle's orbital distance in Section \ref{subsec:smf_study}. For larger values of $f_{sub}$, $\sigma_{T_{diss}}$ can still be quite low for clusters with large orbital distances since the background tidal field does not have a strong radial gradient in the outer regions of the galaxy despite the change in orbital distance due to subhalo interactions being larger. For smaller orbital distances, where the change in orbital distance is smaller for a given $\sigma_e$, $\sigma_{T_{diss}}$ ranges between $3\%$ and $14\%$ depending on the value of $f_{sub}$. The increase in $\sigma_{T_{diss}}$ for clusters with small orbital distances is a result of the tidal field having a strong radial gradient in the inner regions of the galaxy.



\section{Conclusion}\label{sec:conclusion}

Using a large suite of simulations, we explore the effects that $\Lambda$CDM dark matter substructure has on the orbits of massless test particles. 

When evolving test particles in subhalo populations with a single subhalo mass we found that few interactions with higher mass subhalos produce larger orbital deviations than many interactions with low mass subhalos. Subhalos with mass $M_{sh} \geq 10^{8}M_{\odot}$ induce the largest perturbations and deviations in circular orbits. The ineffectiveness of low-mass subhalos at causing test particles to deviate from their initial orbit is likely due to a combination of 1) low mass subhalos not being massive enough to produce strong perturbations and 2) a large number of low-mass subhalo interactions will cancel each other out given that the local population is effectively isotropic.

When evolving test particles in galaxy models where subhalos have a range of masses, we find a lower limit of $f_{sub} = 0.1\%$ for when non-negligible orbital perturbations and deviations will occur. The $f_{sub} = 0.1\%$ simulations contain an average of only 1.2 $\pm$ 0.08 subhalos with masses between $10^{8}M_{\odot}$ and $10^{9}M_{\odot}$. Hence interactions with high mass subhalos will be extremely rare. As a result, potentials with substructure mass fractions equal to and below $0.1\%$ cannot support enough high mass subhalos to yield effective perturbations and orbital deviations in particle orbits. 

In all scenarios we find that perturbations and orbital deviations increase with galactocentric distance. Given that the strength of a galaxy's smooth tidal field will decrease with orbital distance, the relative strength of a given subhalo perturbation will be higher at larger orbital distances. Hence we also find that test particles at larger orbital distances can have their orbits deviate significantly more than particles with smaller orbital distances, despite the lower dark matter density in the outskirts of the galaxy. 

By experimenting with galaxy models with subhalos that are 10 times more dense than predicted by $\Lambda$CDM simulations, we show that denser subhalos produce stronger perturbations to test particles on circular orbits and subsequently larger orbital deviations. Increased subhalo density yield stronger perturbations to test particles because they have a stronger gravitational potential than conventional $\Lambda$CDM subhalos.

Taking into consideration the time dependence of test particle-subhalo interactions, we consider how the range of orbital energies and orbital distances that a given test particle can reach evolve with time. For each simulation, $\sigma_e$ and $\sigma_r$ both increase linearly with time. The rate at which both parameters increases depends on both orbital distance and $f_{sub}$. The linear evolution of $\sigma_e$ is consistent with \citet{Penarrubia2019} (Equations \ref{eq:delE2} and \ref{eq:delv2}), where $\sigma_e(t)$ is derived assuming a single-mass subhalo population with test particles at orbital distances comparable to the mean separation between subhalos (D). To apply Equations \ref{eq:delE2} and \ref{eq:delv2} to our work, we assume the subhalos can be considered to be a single-mass population with a mass equal to the mean sub-halo mass and use the test particle orbital distance to calculate a local value for D. This calculation of a local D effectively takes into consideration both the observed orbital distance and $f_{sub}$ dependence. With these approximations, we find that $\sigma_e(t)$ can be estimated to within 220 $\frac{km}{s^{2}}$. We further fit the evolution of $\sigma_r(t)$ with Equation \ref{eqn:sigr_fit}, where we find the evolution of $\sigma_r$ can be estimated to within 0.08 kpc.

Due to the strong ties between a globular cluster's evolution and its orbital distance, as shown by \citet{baumgardt2003} and \citet{Webb2015}, we explore how a cluster's dissolution time is affected by subhalo induced orbital deviations. We find that, although the relative strength of subhalo perturbations is higher at larger radii and leads to larger orbital deviations, the fractional change in the orbital distance of a test particle is greatest at small radii. Accordingly, we find that clusters that evolve close to the centre of potentials that host populations of substructure have dissolution times that can vary the most from subhalo effects: up to $14\%$ at 5 kpc in potentials with $f_{sub} = 10\%$. Taking into consideration the constraint on the Milky Way's substructure mass fraction from \citet{banik2019}, that out to 20 kpc the substructure mass fraction is proposed to be $0.14^{+0.11}_{-0.07}\%$, we conclude that clusters within 20 kpc of the Galactic centre have not experienced significant orbital deviations from subhalo interactions. Thus clusters within 20kpc of the galactic centre likely also experience negligible change in their dissolution times due to subhalo interactions. With TAP estimating that Milky Way-like galaxies have substructure mass fractions that increase to a maximum of $3\%$ at 100 kpc, clusters on the galactic outskirts could experience subhalo induced orbital deviations on the order of kiloparsecs. Such deviations can lead to outer cluster dissolution times varying by a maximum of $2\%$.

The estimates of the variance of cluster dissolution time presented here are lower limits, as the effects of tidal heating due to non-circular orbits and subhalo interactions are not considered (although it should be noted that \citet{Webb2019} made use of the tidal approximation to determine that subhalo interactions are typically too weak and too short to result in clusters losing additional mass through this mechanism). To simultaneously model the effects of orbital deviations and tidal heating, direct $N$-body simulations of clusters evolving in external tidal fields containing substructure are required. Measurable changes to a cluster's mass or structure can lead to clusters being used to detect the presence of dark matter substructure. If such a method for the indirect detection of subhalos is possible, it would provide estimates for the substructure mass fraction of the Milky Way and test the accuracy of $\Lambda$CDM cosmology.

\section*{Acknowledgements}

NP and JW would like to thank Prof. Roberto Abraham for guidance and feedback through the University of Toronto's Department of Astronomy and Astrophysics undergraduate research program.

\section*{Data Availability}

The data underlying this article will be shared on reasonable request to the corresponding author.




\bibliographystyle{mnras}
\bibliography{example} 










\bsp	
\label{lastpage}
\end{document}